\documentclass[useAMS,usenatbib, usegraphicx]{mn2e}

\def\aj{AJ}%
\def\araa{ARA\&A}%
\def\apj{ApJ}%
\def\apjl{ApJ}%
\def\apjs{ApJS}%
\def\apss{Ap\&SS}%
\def\aap{A\&A}%
\def\mnras{MNRAS}%
\def\pasp{PASP}%
\renewenvironment{thebibliography}[1]{%
 \thebib@list
}{
 \endlist
}

\def\thebib@list{%
 \list{\null}{%
 \partopsep 0mm
 \leftmargin 1.2em
 \labelsep 0mm
 \itemindent -1.2em
 \itemsep 0.0\baselineskip
 \parsep 0mm
  \usecounter{enumi}%
 }%
}%

\title{Host Galaxy Extinction of SNIa: Co-evolution of ISM Structure and Extinction Law with Star-Formation}
\author[B.W. Holwerda]{B. W. Holwerda$^{1}$\thanks{E-mail:holwerda@stsci.edu}\\
$^{1}$Space Telescope Science Institute, 3700 San Martin Drive, Baltimore, MD 21218, USA}

\begin{document}

\date{Accepted ?. Received 29 dec 2007; in original form }

\pagerange{\pageref{firstpage}--\pageref{lastpage}} \pubyear{2007}

\maketitle

\label{firstpage}

\begin{abstract}
This paper presents a mechanism that may modify the extinction law for SNIa observed at higher redshift.
Starting from the observations that (1) SNIa occur predominantly in spiral galaxies, (2) star-formation ejects ISM out of the plane of spirals, (3) star-formation alters the extinction properties of the dust in the ISM, and (4) there is substantially more star-formation at higher redshift, I propose that spiral galaxies have a dustier halo in the past than they do now. The ejected material's lower value of $R_V$ will lead to a lower average value ($\bar{R}_V$) for SNIa observed at higher redshift. 


Two relations in SNIa observations indicate evolution of the average $R_V$: the relation of observed $R_V$ with inclination of the host galaxy at low redshift and the matching of the distribution of extinction values ($A_V$) for SNIa in different redshift intervals. The inclination effect does point to a halo with lower $R_V$ values. In contrast, the distributions of $A_V$ values match best for a $\bar{R}_V(z)$ evolution that mimics the relation of SNIa dimming with redshift attributed to the cosmological constant. However, even in the worse case scenario, the evolution $\bar{R}_V$ can not fully explain the dimming of SNIa: host galaxy extinction law evolution is not a viable alternative to account for the dimming of SNIa.

Future observations of SNIa --multi-color lightcurves and spectra-- will solve separately for values of $A_V$ and $R_V$ for each SNIa . Solving for evolution of $\bar{R}_V$ (and $A_V$) with redshift will be important for the coming generation of cosmological SNIa measurements and has the bonus science of insight into the distribution of dust-rich ISM in the host galaxies in the distant past.

\end{abstract}


\section{\label{sec:intro}Introduction}

Supernova type 1a (SNIa) distance modulus measurements have grown into a powerful measurement of the equation of state of our Universe. Accurate cosmological distances combined with redshift measurements allow for a precise characterisation of the Hubble flow as well as the additional acceleration attributed to the cosmological constant \citep{Riess98,Perlmutter99}. The increasing statistics of SNIa measurements together with more information for each separate supernova event have progressively lowered the observational uncertainties: dust extinction, photometric error and lightcurve characterisation \citep[See e.g.,][]{Tonry03, Knop03, Barris04, Conley06, Astier06}.

Extinction by dust remains a problematic systematic uncertainty in SNIa observations because the applicable extinction law remains poorly understood. This will need to be addressed in order to use SNIa in the next step in accuracy for a cosmological probe.  Dust attenuation affects both the observed SNIa rate and the distance modulus. Extinction by dust occurs in three instances before observation of the SNIa light: 
(1) in our own Milky way, (2) in intergalactic space, and (3) in the host galaxy of the SNIa.

Galactic extinction (1) is a well-studied problem, because it is an ubiquitous one. \cite{Burstein84} produced a  map of Galactic extinction based on HI maps and distant galaxy counts. It was superseded by the map of \cite{Schlegel98}, based on COBE and IRAS far-infrared and sub-mm maps. The latter map is now generally used, together with the Galactic Extinction Law to correct extragalactic sources \citep{Cardelli89, Fitzpatrick99}; the inferred extinction along a line-of-sight is proportional to the reddening: $A_V = R_V \times E(B-V)$. The canonical value of $R_V$ is 3.1 with occasionally lower values towards the Galactic Center \citep{Udalski03} and higher elsewhere \citep{Cardelli89, Fitzpatrick99}.

Extinction by dust in intergalactic space (2) has been proposed as an alternative explanation for SNIa dimming, which is generally attributed to the cosmological acceleration \citep{Aguirre99b, Aguirre99a, Aguirre00}. The resulting extinction law of this dust would effectively be grey\footnote{The term grey extinction is used to denote that there is no relation between the reddening of an object and its extinction.} because the attenuating dust would be spread over all redshifts along every line-of-sight. The coincidence of both uniform distribution in both sky and redshift space does give this explanation a somewhat contrived appearance. The injection of dust into the intergalactic medium would have to be constant and substantial. Models of a dusty universe \citep{Goobar02,Robaina07} find this grey dust explanation increasingly inconsistent with observational data \citep[See also e.g., ][]{Riess07}.

Extinction within the SNIa's host galaxy (3), dust in the immediate surroundings and any disk, ring or spiral arm the line-of-sight passes through in projection, is an observationally evident, yet not fully constrained uncertainty. The Dark Energy Task Group Report \citep{DETF} notes this as a primary source of uncertainty for SNIa measurements.

Three characteristics of the host galaxy's dust could --and are expected to-- change over the history of the universe: (a) total dust content or mass, (b) dust distribution within the host galaxy and (c) dust composition. The overall effect on the effective extinction law is of interest for the distance determination from SNIa light curves.

Dust mass (a) is a variable in several Spectral Energy Distribution (SED) studies of distant galaxies.
\cite{Calzetti99} modeled the overall dust content of galaxies over time and found that a maximum occurred either at z=1 or at z=3. \cite{RR03} modeled the SED's from distant galaxies and similarly found a maximum dust content at z=1. \cite{IP07} find a steady increase of dust mass with time from the UV-IR SED of galaxies. 
The typical dust mass found in distant galaxies is very much a function of the selected sample. Far-infrared selected samples point to dust-rich galaxies, similar to Arp 220 \citep{RR05}, optical/UV selected samples point to disks very similar to the local ones \citep{Sajina06} and Lyman-$\alpha$ galaxies point to low-extinction disks \citep{Nilsson07}. 
However, more dust mass should not affect the SNIa distance determinations if the extinction law remains the same for nearby and distant galaxies.
More dust in the distant galaxies will predominantly affect the observed SNIa {\it rate} \citep{Hatano98, Cappellaro99, Goobar02, Riello05, Mannucci07}, as heavily obscured SNIa drop from observed samples.

The dust distribution in host galaxies (b) is commonly modeled as a double exponential, one radial and one vertical, sometimes with rings to mimic spiral arms. The radial scale of the dust distribution is assumed to be similar to the stellar one and the vertical dust scale is supposed to be much smaller than the stellar one.
Previous observations of the scale-height in nearby (edge-on) galaxies appeared to corroborate the small scale-height \citep[e.g,][]{Xilouris99,Holwerda05b,Bianchi07} but recent observations indicate a much higher scale-height for the dust \citep{Seth05,Kamphuis07}, similar to the stellar scale.
If the average scale-height of the dust distribution was higher in the past, then SNIa in the plane of the disk will encounter more extinction, especially when viewed in an inclined system. The different distribution has a similar effect as variation the dust content of galaxies and only the observed SNIa rate will be affected, unless dust composition and distribution are related. 

The dust composition (c), notably the ratio of small to large grains, directly affects the observed extinction law \citep[See the  review by][]{Draine03}. Evolution in the average extinction law is the most troubling possibility, as this would affect the extinction correction of SNIa and indirectly the measured Hubble flow and acceleration. For local SNIa, variations in the extinction law have been observed \citep[e.g.,][]{Riess96b,Jha07}. \cite{Wang05} explained the different observed extinction law for some SNIa as the effect of circumstellar material around the SNIa progenitor. This is a plausible scenario as the massive-star supernovae are very efficient dust producers \citep{Sugerman06}.  \cite{Patat07} and \cite{Wang07} report observations of such material.

Alternatively, there is substantial evidence for a link between star-formation and extinction law in star-forming galaxies \citep[See the review in][and the references therein.]{Calzetti01}. Star-formation produces more small grains and the intense UV fields alter grain composition.

Since it is not unreasonable to suppose that star-formation affects both the distribution and the extinction characteristic of dust in spiral disks, I propose a simple model for the evolution of the extinction law applicable to SNIa measurements. 
The aim of this paper is to explore this link between the extinction law for SNIa and star-formation of host galaxies and to investigate how much extinction law evolution could reasonably influence SNIa distance measurements. 
The paper is organized in a brief review of current observations of evolution in host galaxy dust and $R_V$ (\S 2), a description of the model (\S 3), two tests based on current SNIa observations (\S 4), and discussion and conclusions (\S 5).

\section{\label{s:status}Observational Status}

Observational evidence for a different extinction law at higher redshifts comes from lensed quasars, QSO reddening by damped Lyman-$\alpha$ systems and GRB afterglows, as well as SNIa measurements. 

Gravitational lenses of QSOs find a range of values for $R_V$ up to a redshift of 1 \citep{Nadeau91, Falco99, Motta02, Toft00, Goicoechea05, Eliasdottir06}. However, the lenses are often elliptical galaxies, which are not typical SNIa hosts.
 \cite{Pei91} find evidence of QSO reddening by Lyman-$\alpha$ systems but this is disputed by \cite{Murphy04}, both based on the SDSS sample. \cite{York06} find a SMC-type extinction based on all SDSS QSOs with a Lyman-$\alpha$ system. \cite{Kann06} compare extinciton for different GRB afterglows and find evidence for dust in the GRB hosts. None of these observations effectively constrain the extinction law applicable to SNIa.

In the SNIa literature, the problem of SNIa host galaxy dust extinction is well recognised \citep{DETF,Conley06,Astier06,Wood-Vasey07}. Observations of high redshift SNIa tell us that their spectra are similar to local ones \citep{Garavini05, Hook05, Garavini07}, with no anomalous reddening \citep{Knop03} and dust masses of host galaxies similar to those of local galaxies \citep[][based on sub-mm data]{Clements04, Clements05}. 

Several studies look at the relations between SNIa lightcurve properties (peak brightness, duration, and colour) and host galaxy properties (type, stellar mass, extinction, and star-formation). 
A relation between SN peak brightness and host galaxy type is well established \citep{Hamuy96a, Hamuy00, Saha97, Saha99, Parodi00, Sandage01, Sullivan03, Reindl05}. 
\cite{Sullivan06} refine the relation to one between the SNIa lightcurve and star-formation in the galaxy: passive galaxies, with no star-formation, preferentially host faster-declining/dimmer SN Ia, while brighter events are found in systems with ongoing star-formation. The SNIa related to ongoing star-formation could well dominate the population at higher redshift \citep{Sullivan06,Mannucci07,Howell07}, as there is much more star-formation.
Host galaxy extinction is of equal importance as any change in the SNIa population. \cite{Sullivan03} report dimmer SNIa for spiral galaxy hosts, especially when projected in or through the inner part of the disk. \cite{Arbutina07} also finds a radial dependence of the extinction of local SNIa. 
Values for $R_V$ different than the canonical $R_V$ = 3.1 have been found for individual SNIa \citep{Riess96b,Krisciunas06,Krisciunas07}. Based on their entire sample, \cite{Reindl05} find a value of 2.65. \cite{Jha07} find a mean value of 2.7 for heavily obscure SN and 2.9 for the less extincted ones. The explanations for deviant $R_V$ values are circumstellar material \cite{Wang07} or a different dust composition \citep{Riess96b,Krisciunas06}. Evolution in $R_V$ may be critical for precision cosmology, and the SNIa themselves are also the best probe to characterize this evolution.

\begin{figure}
\centering
\includegraphics[width=0.45\textwidth]{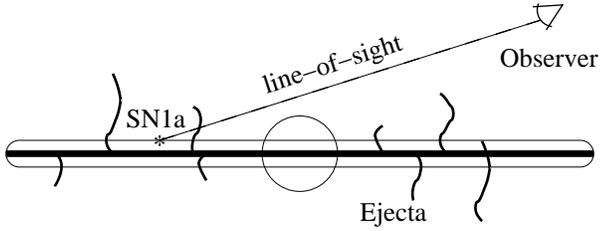}
\caption{A cartoon of our model: star-formation ejects plumes of ISM with lower $R_V$ values out of the disk of the host galaxy. A line-of-sight to a SNIa can intersect such plumes. The higher star-formation at z$\sim$1 results in more plumes and hence a higher probability for a SNIa line-of-sight intersecting one. }
\label{f:cartoon} 
\end{figure}

\section{\label{s:mod}Evolution model of Host Galaxy Extinction}

%
%
Let us consider four observational facts: (I) SNIa occur predominantly in spirals \citep{Sullivan03, Reindl05}, (II) star-formation ejects ISM from the planes of spiral disks and the ejected material takes some time to rain back \citep{Howk97, Howk99b, Howk99a, Dalcanton04, Howk05, Thompson04, Kamphuis07}, (III) star-formation modifies the extinction characteristics \citep[lower $R_V$,][]{Gordon03, Krisciunas06} and (IV) there is an order of magnitude more star-formation at and beyond z=1 \citep{Madau98,Steidel99, Giavalisco04, Thompson06, Hopkins06}. Therefore, at z$\sim$1, spiral disks are likely to have an halo of ISM, recently processed and ejected by star-formation. For the distance measurement with SNIa this is especially important as their light will encounter more $R_V < 3.1$ type dust as a result (See Figure \ref{f:cartoon}). Hence, the {\em average} value of $R_V$ ($\bar{R}_V$) might be lower for SNIa at higher redshift.
The simplest parametrisation of the $\bar{R}_V$ evolution is a second order polynomial:

\begin{equation}
\bar{R}_V =  a  z^2 + b z + 3.1,
\end{equation}

\noindent with the parameters depending on the choice of $\bar{R}_V$ value for the different epochs. 

Of the four assertions above, the relation between star-formation and extinction law (III) is the most tenuous. Evidence for a different extinction law in star-forming galaxies have been found by \cite{Calzetti94}. However, for extended sources such as HII regions, one would need to disentangle the effects of processed dust grains and a clumpy medium \citep[See][]{Natta84}. 

\cite{Gordon03} presents extinction law measurements for single stars, and hence lines-of-sight, in different regions in the SMC and LMC. They find evidence of dust processing by star-formation and lower values of $\bar{R}_V$ in the star-forming parts. Their average $R_V$ values are: SMC Bar; $\bar{R}_V = 2.74$, LMC Supershell; $\bar{R}_V = 2.76$, LMC average sample; $\bar{R}_V = 3.41$. I note that the values for the regions with higher star-formation are lower than the average for the LMC. 

In this paper I explore three models: Model A assumes no evolution in $\bar{R}_V$. 
Model B takes the LMC and SMC values as a template for the galaxies at z=1 and earlier: star-forming galaxies at z$\sim$1 have $\bar{R}_V$ = 2.7 and gas-rich galaxies at z$\sim$2 and beyond have $\bar{R}_V \approx 3.4$. Model C leaves the parameters in equation 1 free to fit the distributions of SNIa extinction values ($A_V$) in a low, intermediate and high redshift sample of SNIa observations.






\section{\label{s:test}Observational tests with SNIa}

There are two suggestive observations of SNIa that support a general model of evolution in host galaxy extinction: (A) the relation between $\bar{R}_V$ for SNIa and host galaxy disk inclination and (B) the distribution of inferred extinction ($A_V$) for a sample of high, intermediate and low redshift SNIa. 

\begin{figure}
\centering
\includegraphics[width=0.45\textwidth]{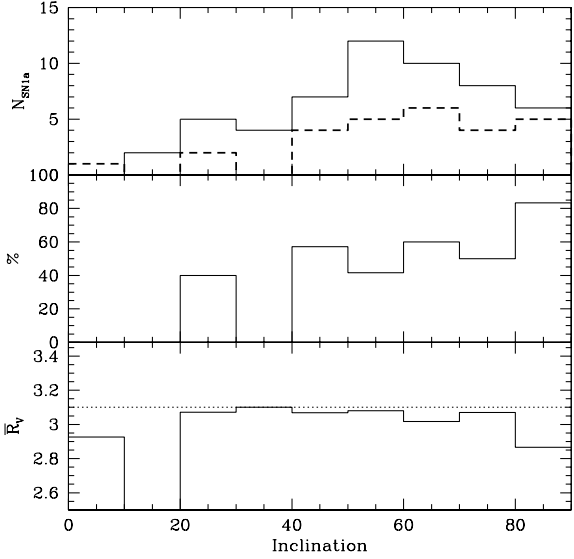}
\caption{The relation between inclination and $R_V$ values from \protect\cite{Jha07}, for the late-type galaxies. Top panel: the number of SNIa with $R_V = 3.1$ and the number with deviant values (dashed line). Middle panel: the percentage of SNIa with deviant $R_V$ values. Bottom panel: the average value of $R_V$. There are very few SN observed in perfectly face-on galaxies. In higher inclined disks the average value of $R_V$ is lower, indicating a possible effect of ejecta in the galaxy's halo.}
\label{f:incl} 
\end{figure}

\cite{Jha07} use the multi-filter lightcurves available for a sample of 133 SNIa to independently fit the value of $R_V$ as well as the light-curve peak and SNIa colour. The inclination and host galaxy type are from {\it Hyperleda}\footnote{http://leda.univ-lyon1.fr/} and $R_V$ values from table 4 in \cite{Jha07}.
Figure \ref{f:incl} shows the distribution of these SNIa as a function of disk inclination of the late-type host galaxies for those SNIa with $R_V=3.1$ (the default) and those with deviant values ($R_V \ne 3.1$). Notably, the deviant values start to dominate at the higher inclinations. Figure \ref{f:incl} also shows the $\bar{R}_V$ value for each inclination bin; it diminishes with increasing inclination. This supports our model of recent ejecta lowering the value of $\bar{R}_V$. 

\begin{figure}
\centering
\includegraphics[width=0.45\textwidth]{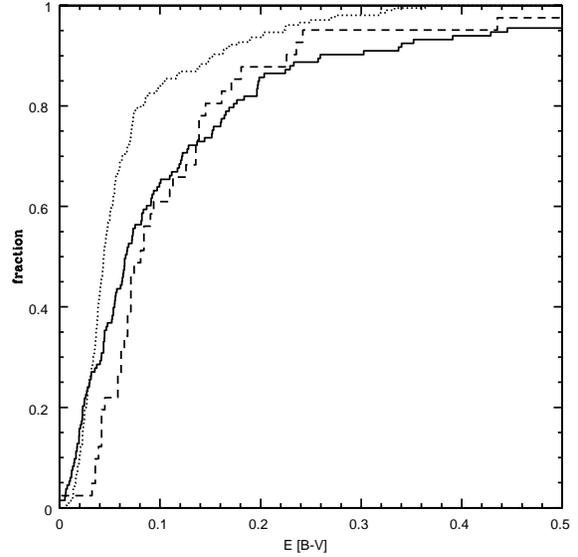}
\caption{\label{f:ebv} The cumulative histogram of SNIa reddening for three samples: the nearby SNIa sample \protect\citep[][solid line]{Jha07}, the intermediate distance sample from the ESSENCE survey \citep[][dotted line]{Wood-Vasey07} and the high redshift sample \protect\citep[][dashed line]{Riess07}. The differences in distribution point to evolution of the extinction in host galaxies, either in $A_V$, $R_V$ or both.}
\end{figure}

The second test is to compare the distributions of $A_V$ of SNIa at different redshift. Figure \ref{f:ebv} shows the distribution of extinction values ($A_V$), based on reddening and a $R_V$ valued of 3.1, for three samples of SNIa, a local, an intermediate and a high-redshift one. Local SNIa are those from the sample of \cite{Jha07}. The intermediate redshift sample is the ESSENCE data from \cite{Wood-Vasey07} and the highest redshift data is from \cite{Riess07}. Sample sizes are very different: 133, 189 and 33 for the low, intermediate, and high redshift samples respectively. The assumption is that the reddening measurements are comparable between these samples. Each of these studies uses a `prior'
distribution for the extinction values to optimize the fit to the lightcurves and I assume here that the affect of this prior is negligible on the reported distribution \citep[See for a good discussion of extinction priors][]{Wood-Vasey07}. 
The distributions of extinction values in Figure \ref{f:ebv} show that there is some type of evolution in $A_V$ from one redshift sample to the next. SNIa searches use a ``gold" and ``silver" standard for SNIa lightcurves. To mimick these I limit the samples to $A_V < 0.25$ and $A_V < 0.5$ respectively in the following analysis.

\begin{figure}
\centering
\includegraphics[width=0.45\textwidth]{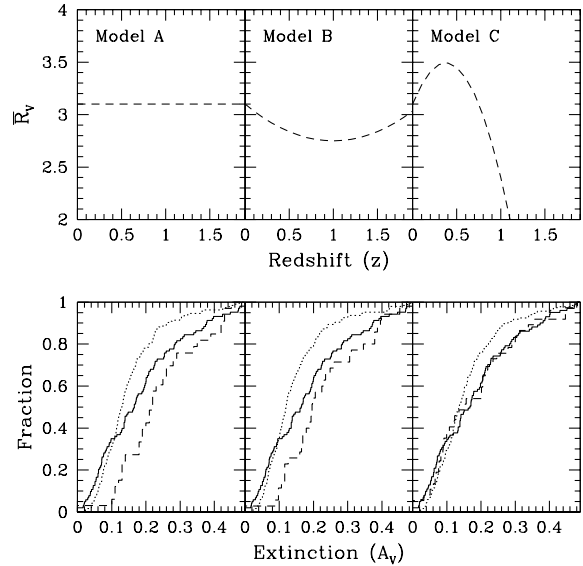}
\caption{\label{f:avmod} Top panels: the three models considered for evolution of  $\bar{R}_V(z)$, (A) no $\bar{R}_V$ evolution, (B) $\bar{R}_V  = a z^2 + b z + 3.1$ with $\bar{R}_V(z=1) = 2.75$ and $\bar{R}_V(z\sim2) = 3.4$, and (C) with $\bar{R}_V  = a z^2 + b z + 3.1$ fit to minimize the difference between the distribution of \protect\cite{Jha07} and the other two distributions ($R_V(z_{max}$ = 0.25 to 0.3) = 3.2 to 4.1).
Bottom panels: the cumulative histogram of $A_V$ of SNIa from a nearby \protect\citep[solid line][]{Jha07}, intermediate \protect\citep[dotted line][]{Wood-Vasey07} and distant sample \protect\citep[dashed line][]{Riess07} for the different models in the top panels). }
\end{figure}

In Figure \ref{f:avmod}, I compare the three different models for $R_V$ evolution (top panels) using the recomputed $A_V$ distribution for each model (bottom panels). I limit the comparison to the SNIa ``silver" standard to maximize statistics. Figure \ref{f:avmod} shows Model A (no evolution, $\bar{R}_V$ = 3.1) and two evolution models (B and C). Model B uses the Magellanic Cloud values from \cite{Gordon03} to parameterise the dependence of $\bar{R}_V$(z). Model C uses the best fit of equation 1 to minimize the differences between the $A_V$ distributions, assuming for each sample the same parent distribution, i.e., random lines of sight through a disk of a spiral galaxy.  

The Kolmogorov-Smirnov probability that all three $A_V$ populations are from the same parent populations without a change in $\bar{R}_V$ (Model A, Figure \ref{f:avmod}, top left panel) is extremely low ($P_{KS} \sim 0$, Figure \ref{f:avmod}, bottom left panel). Hence, some type of evolution is present in the distribution of SNIa extinction values ($A_V$) or extinction law ($\bar{R}_V$).

Model B is what one naively would expect on the basis of the narrative in \S 3: the star-forming galaxies have a lower $\bar{R}_V$, which rises with gas-fraction at higher z and using the \cite{Gordon03} values as a template. The top middle panel in Figure \ref{f:avmod} shows the $\bar{R}_V$ values. I recomputed the values for $A_V$ with the model $\bar{R}_V(z)$ (Figure \ref{f:avmod}, bottom middle panel). This model does not significantly improve the match between the three distributions of SNIa extinctions at different redshifts (see Table \ref{t:mod}). If this model is true, there is substantial independent evolution in the distribution of $A_V$. 

Model C is a fit of the parameters in equation 1 to recompute the $A_V$ distribution and maximizing the probability that they are from the same parent distribution. The best fit values are in Table \ref{t:mod} and Figure \ref{f:avmod} shows the inferred $\bar{R}_V$ (top right) and computed $A_V$ distribution (bottom right). The best model to match the intermediate and high-redshift samples to the low-redshift reference peaks at $z_{max}$ = 0.25 to 0.3 with a high value of $\bar{R}_V(z_{max})$. In this case little or no evolution in $A_V$ is needed. However, the values of $\bar{R}_V$ for each epoch are counter to expectations of \S 3.

\begin{figure}
\centering
\includegraphics[width=0.45\textwidth]{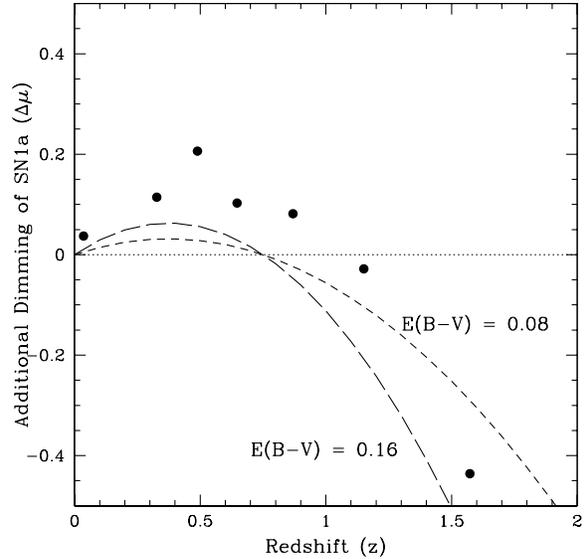}
\caption{\label{f:dm}The maximum possible difference in extinction correction for the ``silver" ($A_V$ = 0.5 for all SNIa) and ``gold" ($A_V$ = 0.25 for all SNIa) samples between $R_V=3.1$ and Model C. The points are the binned SNIa residuals, after subtracting the empty universe model, from \protect\cite{Riess07}, Figure 6. Even if all the SNIa are affected by the maximum $A_V$ {\it and} Model C holds true, evolution in $\bar{R}_V$ {\it alone} cannot explain the residuals in average distance moduli of SNIa.}
\end{figure}

Figure \ref{f:dm} shows the observed average dimming of SNIa as a function of redshift with the 
maximum possible impact of $\bar{R}_V$ evolution plotted: model C with the maximum E(B-V) value for each SNIa allowed by the silver and gold sample selection.
In this worst case scenario I assume that all SNIa are affected by the maximum reddening allowed for selection into the gold or silver samples, even though they are clearly not (Figure \ref{f:ebv}). The worst-case scenario (pure $\bar{R}_V$ evolution) has the same shape as the dimming observed in SNIa, but the effect of $\bar{R}_V$ evolution is not strong enough to fully account for the observed dimming of SNIa. 
The exact origin of $\bar{R}_V$ evolution can be either changes in the immediate surroundings of SNIa or evolution in the ISM of host galaxies.

Because high-redshift observations of SNIa lightcurves are often limited to two filters to minimize observatory time spent on a given object, the applied extinction correction is mostly the fiducial $R_V=3.1$. Occasional evidence from higher-redshift spectra already point to very low values of $R_V$ (A. Riess, private communication). Figure \ref{f:dm} shows that multi-filter and/or spectroscopic determination of $R_V$ for observed SNIa are crucial to accurately determine the equation of state of our Universe from SNIa distances. 

\section{\label{s:disc}Discussion and conclusions}




It seems likely that the distribution of $A_V$ values for SNIa lightcurves will vary as a function of redshift. In this paper, I treated this as evolution of the average extinction law ($\bar{R}_V$). There, however, are three possible explanations for the behavior of the $A_V$ distribution, apart from a fit-prior artifact:
\begin{itemize}
\item[1.] $\bar{R}_V$ does not change much, but due to dust ejection and/or additional dust production (from the elevated star-formation in each galaxy), each SNIa at higher redshift is seen through more host galaxy dust.
\item[2.] The host galaxies are similar to nearby spirals but SNIa occur preferentially in different environments at each redshift. At higher redshifts, they occur in more dust-rich parts of spirals (e.g., the spiral arms or in interacting galaxies).
\item[3] $\bar{R}_V$ --and hence $A_V$-- does evolve as a combination of its dependence on dust composition and dust distribution in the host galaxy. 
\end{itemize}
Options 1 and 2 mean that the observed difference in $A_V$ distribution (Figure \ref{f:avmod}) at different redshifts is real and the inferred distances from SNIa lightcurves suffer no ill effects. In option 3, however, the evolution in the host galaxy extinction law could skew the observations of the acceleration of the expansion of the Universe and the inferred cosmology. 

I would like to stress that option 3 is not as far-fetched as cosmic gray dust and even the worst case scenario would not fully explain the SNIa dimming (Figure \ref{f:dm}).  However, a realistic model would require (1) the relative positions in the disk of SNIa at higher redshifts to characterize the importance of scenario 2, and (2) fits of their lightcurves which include $R_V$ as a fit parameter, similar to those of \cite{Jha07}, to constrain scenario 3. Given enough statistics, the SNIa measurements can then be used as probes of the distribution and severity of host galaxy extinction in disks at high redshift.
In this paper I assumed that the distributions of reddening values are uniformly derived and the choice of prior in the lightcurve analysis did not matter. In future work, one would prefer the same prior in all redshift samples and prefer even more sufficient information to conclusively determine $A_V$ and $R_V$ for each SNIa lightcurve. 

In conclusion:
\begin{itemize}
\item[1] There is a second plausible mechanism that modifies the extinction law for SNIa in higher redshift host galaxies (\S 3); host galaxy dust evolution in addition to material around the SNIa progenitor.
\item[2] The toy model of this mechanism does not match the $A_V$ distribution of SNIa samples at different redshifts (\S 4, Figure \ref{f:avmod}, Model B).
\item[3] Matching up $A_V$ distributions leads to a $\bar{R}_V$ evolution model that resembles the shape and peak of the observed average dimming of SNIa (\S 4, Figure \ref{f:avmod}, Model C).
\item[4] Taken as a worst-case scenario, this model still cannot account for the observed average dimming of SNIa with redshift. 
\item[5] It is likely that there is evolution in the distribution of $A_V$ values with redshift, making it impossible to deduce $\bar{R}_V$ evolution from these distributions alone. As both affect the precision cosmology to be done with the next generation of SNIa measurements, it is paramount that new SNIa observations solve for both $A_V$ and $R_V$ separately.
\item[6] The bonus science of measuring both $A_V$ and $R_V$ for each SNIa is a probe in the distribution, content and composition of dust in their host galaxies.
\end{itemize}


\section*{Acknowledgments}

The author would like to thank Michael Wood-Vasey, Andy Fruchter, Rosa Gonz\'{a}les, Daniela Calzetti, Ron Allen and Adam Riess for useful discussions and comments on SNIa host galaxy extinction. 

\begin{table}
\caption{\label{t:mod}The values for two different toy models, the peak redshift and peak value of $\bar{R}_V$ and the Kolmogorov-Smirnov test }
\begin{center}
\begin{tabular}{l | lll lll}

			& \multicolumn{3}{l}{ESSENCE}			& \multicolumn{3}{l}{\cite{Riess07}}  \\
			& $z_{max}$ & $R_{max}$ & $P_{KS}$	& $z_{max}$ & $R_{max}$ & $P_{KS}$	\\
\hline
\hline
gold			& 		& 		& 				& 		& 		&  \\
($A_V < 0.25$)	& 		& 		& 				& 		& 		&  \\
Model B		& 1		& 2.75	& 0.3				& 1		& 2.75	& 0.5	\\			
Model C		& 0.3		& 4.1		& 72				& 0.25	& 3.2		& 97\\
\hline
silver		& 		& 		& 				& 		& 		&  \\
($A_V < 0.5$)	& 		& 		& 				& 		& 		&  \\
Model B		& 1		& 2.75	& 0.2				& 1		& 2.75	& 2\\
Model C		& 0.4		& 4.3		& 33				& 0.4		& 3.4		& 99 \\
\hline
\end{tabular}
\end{center}
\end{table}%

\section*{References}


\end{document}